\documentclass[twocolumn]{autart}

\usepackage{amsmath}
\usepackage{amssymb}
\usepackage{enumerate}
\usepackage{graphicx}
\usepackage{xcolor}
\usepackage[round]{natbib}
\usepackage{times}
\usepackage{bm}
\usepackage{multirow}
\usepackage{diagbox}

\def\RR{{\mathbb R}}

\def\sign{\,\mathrm{sign}}
\def\comp{\,{\scriptstyle\circ}\,}
\def\QED{$\hfill\blacksquare$}

\newcommand{\abs}[1]{\left|#1\right|}
\newcommand{\norm}[1]{\left\|#1\right\|}
\newcommand{\spow}[2]{\left\lfloor#1\right\rceil^{#2}}
\newcommand{\spowf}[3]{\spow{#1}{\frac{#2}{#3}}}

\newcommand{\diffd}{\mathrm{d}}

\newcommand{\deriv}[2]{\frac{\diffd {#1}}{\diffd  {#2}}}

\newcommand{\pderiv}[2]{\frac{\partial{#1}}{\partial{#2}}}

\newcommand{\sbrm}[1]{\sb{\mathrm{#1}}}

\newcommand{\TT}{^{\mathrm{T}}}

\newcommand{\x}{\bm{x}}
\newcommand{\D}{\bm{D}}
\newcommand{\ka}{\bm{k}}
\newcommand{\z}{\bm{z}}
\renewcommand{\v}{\bm{v}}
\newcommand{\A}{\bm{A}}

\newcommand{\g}{\bm{g}}
\renewcommand{\e}{\bm{e}}
\newcommand{\ee}{\mathrm{e}}

\newcommand{\Tb}{\hat{\mathbb{T}}}

\newtheorem{lemma}{Lemma}[section]
\newtheorem{theorem}[lemma]{Theorem}
\newtheorem{defin}{Definition}
\newtheorem{prp}[lemma]{Proposition}
\newtheorem{remark}[lemma]{Remark}

\newtheorem{prb}{Problem}

\renewcommand{\log}{\ln}

\begin{document}

\begin{frontmatter}
    \title{Robust Exact Differentiators with Predefined Convergence Time\thanksref{footnoteinfo}} 

\thanks[footnoteinfo]{Work partially supported by the Christian Doppler Research Association, the Austrian Federal Ministry for Digital and Economic Affairs and the National Foundation for Research, Technology and Development, by CONACYT (Consejo Nacional de Ciencia y Tecnología) grant 282013, by PAPIIT-UNAM (Programa de Apoyo a Proyectos de Investigación e Innovación Tecnológica) grant IN115419 and by ANPCyT grant PICT 2018-1385, Argentina.}

\author[Seeber]{Richard Seeber}\ead{richard.seeber@tugraz.at},
\author[Haimo]{Hernan Haimovich}\ead{haimovich@cifasis-conicet.gov.ar},
\author[Seeber]{Martin Horn}\ead{martin.horn@tugraz.at},
\author[Fridman]{Leonid Fridman}\ead{lfridman@unam.mx},
\author[DeBa]{Hern\'an De Battista}\ead{deba@ing.unlp.edu.ar}

\address[Seeber]{Christian Doppler Laboratory for Model Based Control of Complex Test Bed Systems, Institute of Automation and Control, Graz University of Technology, Graz, Austria}
\address[Haimo]{Centro Internacional Franco-Argentino de Ciencias de la Informaci\'on y de Sistemas (CIFASIS), CONICET-UNR,\\ Ocampo y Esmeralda, 2000 Rosario, Argentina.}
\address[Fridman]{Engineering Faculty, National Autonomous University of Mexico (UNAM), Ciudad Universitaria, Mexico}
\address[DeBa]{Grupo de Control Aplicado (GCA), Instituto LEICI, UNLP-CONICET, La Plata, Argentina.}


\begin{abstract}
    The problem of exactly differentiating a signal with bounded second derivative is considered.
    A class of differentiators is proposed, which converge to the derivative of such a signal within a fixed, i.e., a finite and uniformly bounded convergence time.
    A tuning procedure is derived that allows to assign an arbitrary, predefined upper bound for this convergence time.
    It is furthermore shown that this bound can be made arbitrarily tight by appropriate tuning.
    The usefulness of the procedure is demonstrated by applying it to the well-known uniform robust exact differentiator, which is included in the considered class of differentiators as a special case.
\end{abstract}

\end{frontmatter}

\section{Introduction}
\label{sec:intro}

The real-time computation of the exact derivative of a measured signal is a very important problem that has many useful applications, such as robust state estimation and fault detection.
A variety of successful approaches to real-time signal differentiation is based on homogeneous sliding-mode algorithms, see, e.g. \citep{levant_auto98,levant_ijc03,levliv_ifacol17}.
Due to homogeneity, these approaches have several useful features: robustness with respect to measurement noise, convergence to the derivative in finite time, and existence of appropriate discrete-time implementations \citep{livlev_auto14,koch2019discreteequivalents,koch2020discrete}.

Among these features, \emph{finite-time convergence} \citep[see][]{roxin1966finite,bhaber_siamjco00} is one of the most important advantages because, in theory, it allows to exactly reconstruct states and even perturbations within a finite time \citep{flobar_ijss07,bejarano2007hierarchical,li2014disturbance}.
In a control context, this advantage permits to separate design and stability analysis of the differentiator (i.e., the observer) from that of a nonlinear state-feedback controller.
This separation becomes possible because the controller can be switched on after the differentiator has converged \citep{angulo2013output}.
In practice, an upper bound for the finite convergence time is required.
While such a bound always exists, it may grow infinitely large with increasingly large initial conditions.
To overcome this disadvantage, a stronger stability notion was introduced by \citet{crumor_tac11} \citep[see also][]{angmor_auto13} that is uniform with respect to the initial condition and was later called \emph{fixed-time convergence} \citep{polyakov2011nonlinear,polyakov2014stability}.
This property guarantees that the convergence time is uniformly upper bounded by a fixed finite time irrespective of the initial condition.
While fixed-time convergent differentiators are not homogeneous, they are typically designed to retain most of the corresponding useful features.

One of the most successful approaches for differentiation in \emph{finite time} is the first-order robust exact differentiator \citep{levant_auto98}.
This differentiator is based on the so-called super-twisting algorithm (STA) originally proposed by \citet{levant1993sliding}, and can differentiate signals with Lipschitz continuous time derivative.
Higher-order derivatives can be obtained using its arbitrary-order generalization \citep{levant_ijc03} or also by cascading first-order differentiators in a step-by-step manner \citep{flobar_ijss07,bejarano2007hierarchical}. 

Several extensions of STA-based differentiators with \emph{fixed-time} convergence exist.
Examples are the uniformly convergent arbitrary order differentiator \citep{angmor_auto13} and the first-order uniform robust exact differentiator proposed by \cite{crumor_tac11}.
For the latter, a generalization was presented by \citet{moreno2011lyapunov} in the form of the generic second order algorithm, which itself is further generalized as the disturbance-tailored STA by \citet{haideb_auto19}.

An important practical problem is how to make the convergence time satisfy a desired global upper bound.
Together with state-feedback control, this allows the designer to tune beforehand also the time after which the controller is turned on.
For the uniform robust exact differentiator, this problem is studied empirically in \citet{fraguela2012design}.
More recently, the related concepts of \emph{prescribed-time stability} by \cite{holloway2019prescribed} and \emph{predefined-time stability} by \cite{sanchez2015predefined,sangom_jmci17} have been proposed in that regard.
The former concept studies ways to prescribe the actual convergence time, rather than its upper bounds, typically by employing time-varying gains that tend to infinity at the desired convergence time instant \citep{holloway2019prescribed} or by varying the homogeneity degree \citep{chitour2020stabilization}.
The latter concept studies systems for which an arbitrary convergence time bound can be specified by a suitable choice of some free parameters \citep{sanchez2015predefined} or by means of a time-varying redesign of given fixed-time stable systems \citep{gomez2020design}.
Algorithms studied in that regard typically are focused on closed loops obtained from controllers, however, rather than differentiators.

Causing a sliding-mode differentiator to satisfy a prescribed convergence time is a challenging problem that requires reasonably tight convergence time bounds.
For fixed-time differentiators based on the STA, convergence time bounds were obtained in the course of a Lyapunov-based stability analysis \citep[see, e.g.][]{moreno2011lyapunov}.
Simple convergence time bounds for another differentiator approach are given by \citet{basin2019finite} (see also references therein). This approach, however, is suitable only for functions with constant time derivative, as shown by \citet{seeber2020three}.
For the STA, i.e., the first-order robust exact differentiator, convergence time bounds have been studied more extensively, see, e.g., the recent tutorial \citep{seeber2020computing} and references therein.
Notably, asymptotically exact bounds have been obtained in \cite{seehor_tac18} by computing the STA's convergence time in the form of an improper integral.

The present paper proposes a new class of first-order fixed-time convergent differentiators along with a tuning paradigm for assigning an upper bound to its global uniform convergence time.
The considered differentiator is based on the disturbance-tailored STA from \citet{haideb_auto19}, and thus includes the uniform robust exact differentiator as a special case.
The differentiator's convergence time is studied by extending the convergence time computation and the corresponding asymptotically exact bounds from \citet{seehor_tac18}.
The tuning paradigm is obtained by combining these bounds with a homogeneity-like scaling property of the global convergence time.

The paper is structured as follows.
Section~\ref{sec:prob-stat} states the considered problem of obtaining a signal's time derivative within a given finite time, and Section~\ref{sec:prop-diff-struct} discusses the differentiator structure used for this purpose and its properties.
Section~\ref{sec:design-proc} proposes a tuning procedure to solve the problem.
The main \mbox{theorem---Theorem~\ref{th:main}---permits} to specify a bound for the differentiator's convergence time based on the computation of only a single convergence time bound for an arbitrary parameter setting.
After illustrating the proposed tuning procedure in Section~\ref{sec:design-example} by means of examples, Sections~\ref{sec:conv-time-func} and~\ref{sec:global-conv-time} compute the convergence time function and study the global convergence time of the proposed differentiator, thereby providing the tools required for the presented tuning paradigm.
Conclusions are given in Section~\ref{sec:conclusions}.
Section~\ref{sec:proofs} forms an appendix that provides proofs for the technical lemmas used in the paper.

\textbf{Notation:}
$\RR$, $\RR_{\ge 0}$, $\RR_{>0}$ denote the reals, nonnegative reals, and positive reals, respectively.
If $\alpha\in\RR$, then $|\alpha|$ denotes its absolute value.
For $x\in\RR^n$, $\norm{x}$ denotes its Euclidean norm.
For $y, p \in \RR$, the abbreviations $\spow{y}{p} = \abs{y}^p \sign(y)$ and $\spow{y}{0} = \sign(y)$ are used.
The symbol `$\comp$' denotes function composition.
If $\nu : D\subseteq\RR \to \RR$, then $\nu'$ denotes the derivative of $\nu$.
A superscript $\TT$ denotes matrix or vector transposition.
For a function $\gamma : \RR_{>0} \to \RR$, define $\gamma(0):= \lim_{s\to 0^+} \gamma(s)$ provided the limit exists.
Furthermore, $\gamma^{-1}$ denotes the inverse function, if it exists, and $\gamma'$ is written for the derivative of the function with respect to its argument, unless it is a function of time, in which case $\dot \gamma$ denotes its time derivative.
The convention $\inf A = \infty$ is used if $A$ is the empty set.
$\lambda_{\min}(\cdot)$ and $\lambda_{\max}(\cdot)$ denote the minimum and maximum eigenvalues, respectively, of a symmetric matrix, and $\operatorname{diag}(d_1, \ldots, d_n) \in \RR^{n \times n}$ denotes a diagonal matrix with diagonal entries $d_1, \ldots, d_n$.

\section{Problem Statement}
\label{sec:prob-stat}

Let $f : \RR_{\ge 0} \to \RR$ be a continuously differentiable function, whose time derivative $\dot{f}$ is globally Lipschitz continuous, i.e., whose second time derivative $\ddot{f}$ for almost all $t$ satisfies
\begin{equation}
    \label{eq:ddotfmax}
    \abs{\ddot{f}(t)} \le L
\end{equation}
with a nonnegative Lipschitz constant $L$.
The goal is to exactly reconstruct the time derivative $\dot f$ within a desired, fixed time $T>0$, i.e., to obtain $\dot{f}(t)$ for $t \ge T$ from $f(t)$ for $t \ge 0$.
Thus, the following problem is addressed.
\begin{prb}
    \label{prb:diff}
  Given a Lipschitz constant $L\ge 0$ and a time $T>0$, construct
  a dynamic system---a differentiator---that takes a function $f$
  as input and generates an output $y$ satisfying $y(t) = \dot{f}(t)$
  for all $t\ge T$, for every function $f$ satisfying (\ref{eq:ddotfmax}).
\end{prb}

\section{Differentiator Structure}
\label{sec:prop-diff-struct}

Problem~\ref{prb:diff} will be solved employing a differentiator having the structure
\begin{subequations}
    \label{eq:diff}
    \begin{align}
      \dot y_1 &= k_1 \nu_1(f - y_1) + y_2 \\
      \label{eq:diff:y2}
      \dot y_2 &= k_2 \nu_2(f - y_1)\\
      y &= y_2
    \end{align}
\end{subequations}
with state variables $y_1, y_2$, functions $\nu_1, \nu_2 : \RR \to \RR$ given by
\begin{align}
    \label{eq:def:nu1nu2}
    \nu_1(x) &= k_3^{-1} \Phi(k_3^{2} x), &
    \nu_2(x) &= 2 \Phi(k_3^{2} x) \Phi'(k_3^{2} x),
\end{align}
and parameters $k_1, k_2, k_3 > 0$.
The function $\Phi : \RR \to \RR$ is called the Differentiator Generating Function (DGF) and is defined next.
Afterwards, the convergence time function of the differentiator and its uniform upper bound, the global convergence time, are introduced.
The differentiator's asymptotic robustness properties are also discussed.

\subsection{Differentiator Generating Function}

The DGF $\Phi$ has to satisfy the following definition.
\begin{defin}[Differentiator Generating Function]
    \label{def:dgf}
    A locally absolutely continuous function $\Phi : \RR \to \RR$ is called a Differentiator Generating Function (DGF) if it has the following properties:
    \begin{enumerate}[(i)]
        \item
            \label{it:Phi:odd}
            $\Phi$ is odd, i.e., $\Phi(-x) = - \Phi(x)$ for all $x \in \RR$,
        \item
            \label{it:Phi:differentiable}
            $\Phi$ and $\Phi'$ are continuously differentiable in $\RR\setminus\{0\}$,
        \item
            \label{it:Phi:increasing}
            $\Phi'$ is positive, i.e., $\Phi'(x) > 0$ for all $x \in \RR \setminus\{0\}$,
        \item
            \label{it:Phi:limit}
            $\lim_{x \to 0} |\Phi'(x)| = \infty$,
        \item
            \label{it:Phi:discont}
$\lim_{x \to 0} \abs{\frac{2 \Phi'(x)^3}{\Phi''(x)}} = 1$.
    \end{enumerate}
\end{defin}

According to conditions \eqref{it:Phi:odd} and \eqref{it:Phi:differentiable}, $\nu_1$ and $\nu_2$ both are odd functions that are continuous and locally Lipschitz in $\RR \setminus\{0\}$. 
The remaining conditions---in particular the limits---yield the following properties of the functions $\nu_1$ and $\nu_2$, which are formally proven in Section~\ref{sec:proofs}.
\begin{lemma}
    \label{lem:nu1nu2:limits}
    Consider a DGF $\Phi$ and let $k_3 > 0$.
    Then, the functions $\nu_1, \nu_2$ defined in \eqref{eq:def:nu1nu2} satisfy
    \begin{equation}
        \label{eq:nu1nu2:limits}
        \lim_{\alpha \to 0} \frac{\nu_1(\alpha^2 x)^2}{\alpha^2} = \abs{x}, \qquad
        \lim_{\alpha \to 0} \abs{\nu_2(\alpha^2 x)} = 1
    \end{equation}
    uniformly in $x$ on every compact subset of $\RR \setminus \{0\}$.
\end{lemma}

As a consequence of \eqref{eq:nu1nu2:limits}, and since the function $\nu_2$ is odd, it is discontinuous in the origin.
    Hence, the right-hand side of \eqref{eq:diff:y2} is discontinuous, and solutions of \eqref{eq:diff} are understood in the sense of \citet{filippov_book88}.
When $y_1 = f$, specifically, \eqref{eq:diff:y2} is to be read as the differential inclusion
\begin{equation}
    \dot y_2 \in \bigl[-k_2 \lim_{x \to 0^{+}} \nu_2(x), -k_2 \lim_{x \to 0^{-}} \nu_2(x)\bigr] = [-k_2, k_2].
\end{equation}
From this relation, one can see that $L \le k_2$ is a necessary condition for the differentiator to converge, because maintaining $y_2(t) = \dot f(t)$ requires $|\ddot f(t)| \le k_2$.

\begin{remark}
    Functions that satisfy the conditions of Definition~\ref{def:dgf} are $\Phi(x) = \spowf{x}{1}{2}$ and $\Phi(x) = \spowf{x}{1}{2} + \spowf{x}{3}{2}$, for example.
    These yield the robust exact differentiator proposed in \cite{levant_auto98} and the uniform robust exact differentiator proposed in \cite{crumor_tac11}, respectively.
\end{remark}

By means of the selected structure, the more specific problem that will be solved can now be stated.

\begin{prb}
  \label{prb:main2}
  Given $L\ge 0$ and $T>0$, select a suitable DGF $\Phi$ and positive design parameters $k_1,k_2,k_3$ such that solutions of \eqref{eq:diff} satisfy $y(t) = \dot{f}(t)$ for all $t\ge T$, for every function $f$ satisfying (\ref{eq:ddotfmax}) and every initial condition $y_1(0),y_2(0)\in\RR$.
\end{prb}

\subsection{Differentiator Error and Convergence Time}
\label{sec:conv-time-defs}

To study the differentiator's convergence behavior, the error variables $x_1 := f - y_1$, $x_2 := \dot{f} - y_2$ are introduced and aggregated in the vector $\x = [ x_1 \quad x_2 ]\TT$ in the following.
For notational convenience, the family of functions $\Phi_{\epsilon}: \RR \to \RR$ is furthermore defined as
\begin{equation}
    \label{eq:Phieps}
    \Phi_\epsilon(x) = \epsilon^{-1} \Phi(\epsilon^{2} x),
\end{equation}
which satisfies $\Phi'_\epsilon(x) = \epsilon \Phi(\epsilon^2 x)$.
Using this notation, one has $\nu_1 = \Phi_{k_3}$ and $\nu_2 = 2\Phi_{k_3} \Phi_{k_3}'$.
The corresponding error dynamics then are
\begin{subequations}
    \label{eq:diffe}
    \begin{align}
    \label{eq:diffe:x1}
    \dot x_1 &= -k_1 \nu_1(x_1) + x_2 = -k_1 \Phi_{k_3}(x_1) + x_2 \\
    \label{eq:diffe:x2}
\dot x_2 &= -k_2 \nu_2(x_1) + \ddot f = - 2 k_2 \Phi_{k_3}(x_1) \Phi_{k_3}'( x_1) + \ddot f
    \end{align}
    with
    \begin{equation}
    \label{eq:diffe:fddot}
        \abs{\ddot f} \le L.
    \end{equation}
\end{subequations}
One can see that $\ddot f$ enters this system as a perturbation.
The case $\ddot f = 0$, i.e., $L = 0$, is hence called \emph{unperturbed case} in the following, while $L \ge 0$ is referred to as \emph{perturbed case}.

If $\Phi$ satisfies Definition~\ref{def:dgf}, then \eqref{eq:diffe} is a well-defined Filippov inclusion, whose solutions exist and are continuable in forward time.
Let $\x(\cdot,\x_0,f)$ denote the solution to (\ref{eq:diffe}) satisfying $\x(0,\x_0,f) = \x_0$. 
Let $\tau(\x_0,f)$ denote the minimum time $t$ that the trajectory $\x(\cdot,\x_0,f)$ takes to converge to the origin, i.e.,
\begin{align}
  \label{eq:deftau}
  \tau(\x_0,f) &= \inf \{t\ge 0 : \x(\sigma,\x_0,f) = 0\quad \forall \sigma \ge t\}.
\end{align}
Depending on the DGF $\Phi$, the parameters
\begin{equation}
  \label{eq:defka}
  \ka := (k_1,k_2,k_3),
\end{equation}
the initial state $\x_0$ and the function $f$, the convergence time $\tau(\x_0,f)$ may be finite or infinite. 

The worst-case convergence time obtained for any $f$ satisfying \eqref{eq:ddotfmax} as a function of the initial state $\x_0$ is called the \emph{convergence time function} of system \eqref{eq:diffe}.
It is denoted by $T^{\Phi,\ka}_L : \RR^2 \to \RR$ and is given by
\begin{equation}
    T^{\Phi,\ka}_L(\x_0) := \sup_{\abs{\ddot f} \le L} \tau(\x_0, f).
\end{equation}
The smallest uniform upper bound of this function with respect to the initial state
\begin{equation}
    \label{eq:defTb}
    \Tb(\Phi, \ka, L) := \sup_{\x_0 \in \RR^2} T^{\Phi, \ka}_L(\x_0)
\end{equation}
is called the differentiator's \emph{global convergence time}.
Using this notation, Problem~\ref{prb:main2} may be restated as follows.

\setcounter{prb}{1}
\begin{prb}[restated]
  \label{prb:main}
  Given $L\ge 0$ and $T>0$, select a suitable DGF $\Phi$ and positive design parameters $k_1,k_2,k_3$ so that the differentiator's global convergence time satisfies $\Tb(\Phi, \ka, L) \le T$. 
\end{prb}
\renewcommand{\theprb}{\arabic{prb}}

Note that, depending on $\Phi$ and $\ka$, the global convergence time may be finite or infinite even when the convergence time function yields finite values for all initial states.
Fixed-time convergence refers to the case of finite $\Tb(\Phi,\ka,L)$, whereas (global) finite-time convergence just indicates that $T^{\Phi,\ka}_L(\x_0)$ is finite for every $\x_0 \in \RR^2$ but does not guarantee that $\Tb(\Phi,\ka,L)$ is also finite. 
For these cases, Lyapunov stability of the origin will additionally be shown, which guarantees that solutions of \eqref{eq:diffe} are also unique in forward time.

\subsection{Asymptotic Properties and Robustness}

Homogeneously approximating the inclusion \eqref{eq:diffe} at the origin in the sense of \cite[Definition 2.1]{andrieu2008homogeneous} with degree $-1$ and weights $(2,1)$ yields the super-twisting algorithm, i.e., the first-order robust exact differentiator:
\begin{theorem}[Asymptotic Behavior]
    \label{th:asymptotic}
    For every DGF $\Phi$ and every $k_3 > 0$, the STA system
    \begin{equation}
        \label{eq:diffe:approx}
        \dot x_1 = -k_1 \spowf{x_1}{1}{2} + x_2, \qquad
        \dot x_2 = -k_2 \spow{x_1}{0} + \ddot f
    \end{equation}
    is a homogeneous approximation\footnote{Note that \cite[Definition 2.1]{andrieu2008homogeneous} is applicable only to autonomous vectorfields (i.e., for $\ddot f = 0$ here), but its extension to the non-autonomous case considered here is straightforward, provided that uniformity with respect to the input is also imposed.} of \eqref{eq:diffe} at the origin.
\end{theorem}
\begin{remark}
    As a consequence of this theorem, asymptotic (small-signal) measurement noise gains and robustness properties of the proposed differentiator are the same as for the first-order robust exact differentiator, see \cite{levant_auto98}, provided that $k_1, k_2$ are chosen such that \eqref{eq:diffe} and \eqref{eq:diffe:approx} both converge in finite time for all $\ddot f$ satisfying \eqref{eq:diffe:fddot}.
\end{remark}
\begin{pf}
Using the limits \eqref{eq:nu1nu2:limits} and symmetry properties of $\nu_1, \nu_2$ defined in \eqref{eq:def:nu1nu2}, one obtains from \eqref{eq:diffe}
    \begin{align*}
\lim_{\alpha \to 0} \frac{-k_1 \nu_1(\alpha^2 x_1) + \alpha x_2}{\alpha} &= - k_1 \spowf{x_1}{1}{2} + x_2, \nonumber \\
        \lim_{\alpha \to 0} \bigl( - k_2 \nu_2(\alpha^2 x_1) + \ddot f \bigr) &= - k_2 \spow{x_1}{0} + \ddot f.
    \end{align*}
Due to Lemma~\ref{lem:nu1nu2:limits}, the limits are uniform in $x_1, x_2$ on compact sets; hence, \eqref{eq:diffe:approx} is a homogeneous approximation of \eqref{eq:diffe} according to \cite[Definition 2.1]{andrieu2008homogeneous}.
    \QED
\end{pf}

\section{Design Procedure}
\label{sec:design-proc}

This section provides a solution to Problem~\ref{prb:main2}.
First, requirements on the DGF $\Phi$ are formulated, which provide a guideline for its design.
Then, a tuning procedure for the parameters $k_1, k_2, k_3$ is proposed, which is based on a \emph{single} upper bound of the global convergence time for \emph{one} set of parameter values.
Thus, a complete design procedure is developed.

\subsection{Design of the Differentiator Generating Function}
\label{sec:dgf-design}

Selecting a DGF according to Definition~\ref{def:dgf} provides for a well-posed Filippov inclusion and will be shown in Section~\ref{sec:conv-time-func} to yield a finite convergence time $T_0^{\Phi,\ka}(\x_0)$ for $L = 0$.
This does not guarantee, however, that the global convergence time $\Tb(\Phi, \ka, L)$ is finite, especially for $L > 0$.
In the following, conditions on the DGF are therefore formulated that will allow to establish a finite global convergence time bound.
They are motivated by the following result, which is proven in Section~\ref{sec:global:lower}.
\begin{prp}
    \label{prop:motivation}
    Let $\ka = (k_1,k_2,k_3) \in \RR_{>0}^3$, $L \in \RR_{\ge 0}$ and consider a DGF $\Phi$.
    If $k_1^2 \ge 8 k_2$, then
    \begin{equation}
        \label{eq:lowerbound}
\Tb(\Phi, \ka, L) \ge \frac{2}{\bigl( k_1 - \sqrt{k_1^2 - 8 k_2} \bigr) k_3} \int_{0}^{\infty} \frac{1}{\Phi(x)} \, \diffd x.
    \end{equation}
\end{prp}

One can see that under the conditions of this proposition, the DGF $\Phi$ has to be selected such that the integral on the right-hand side of \eqref{eq:lowerbound} exists and is finite.
The concept of an admissible DGF is thus introduced, which satisfies this condition along with some additional technical assumptions.
\begin{defin}[Admissibility of DGFs]
    \label{def:dgf:admissible}
    A DGF $\Phi$ is called admissible if there exist constants $B > 0$, $C > 0$, and $D \ge 1$  such that $\Phi$ satisfies
    \begin{enumerate}[(i)]
        \item
            \label{it:Phi:integral}
$\int_{0}^{\infty} \frac{1}{B \Phi(x)} \, \diffd x \le 1$,
        \item
            \label{it:Phi:minimum}
            $C \Phi'(x) \ge 1$ for all $x \in \RR \setminus \{ 0 \}$,
        \item
            \label{it:Phi:growth}
            $2 D |\Phi'(x)|^3 \ge |\Phi''(x)|$ for all $x\in\RR\setminus \{ 0 \}$.
\end{enumerate}
\end{defin}
Condition \eqref{it:Phi:integral} is required for fixed-time convergence, i.e., for the global convergence time to be finite.
The other two conditions \eqref{it:Phi:minimum} and \eqref{it:Phi:growth} can be interpreted as a uniform variant of item \eqref{it:Phi:increasing} and a global variant of item \eqref{it:Phi:discont} of Definition~\ref{def:dgf}, respectively.
They will be used to establish a convergence time bound when $L > 0$.
In particular, condition \eqref{it:Phi:minimum} guarantees that $\Phi(x)$ grows without bound as $x \to \infty$;
and condition \eqref{it:Phi:growth}  implies the existence of a uniform lower bound for $\abs{2 \Phi_{k_3}(x) \Phi_{k_3}'(x)}$, because it is equivalent to
\begin{equation}
    \deriv{\Phi(x)}{x} \ge \frac{1}{2 D} \deriv{}{x} \frac{1}{\Phi'(x)},
\end{equation}
i.e., $\Phi$ grows faster than $\Phi'$ decays.
Such a lower bound is necessary for finite-time stability with $L > 0$, because otherwise system \eqref{eq:diffe} exhibits an additional non-zero equilibrium for any arbitrarily small (and constant) $\ddot f$.

\begin{remark}
    Both $\Phi(x) = \spowf{x}{1}{2}$ and $\Phi(x) = \spowf{x}{1}{2} + \spowf{x}{3}{2}$ satisfy item \eqref{it:Phi:growth} of Definition~\ref{def:dgf:admissible} with $D = 1$.
    Only the latter is actually admissible, however, due to item~\eqref{it:Phi:integral}.
\end{remark}
\begin{remark}
    Note that in the context of the disturbance-tailored super-twisting (DTST) algorithm proposed by \cite{haideb_auto19}, the pair $(\nu_1, \nu_2)$ defined in \eqref{eq:def:nu1nu2} is DTST-admissible when $\Phi$ is an admissible DGF.
\end{remark}

\subsection{Tuning of the Differentiator's Parameters}
\label{sec:tuning}

The proposed tuning procedure, i.e., the procedure for selecting appropriate values for the differentiator's parameters $k_1, k_2, k_3$, is based on the following proposition, which in part has also been noted empirically by \cite{fraguela2012design}.
A formal proof is provided in Section~\ref{sec:global:scaling}.
\begin{prp}[Scaling of Parameters]
  \label{prop:tuning}
  Let $\ka \in \RR_{>0}^3$, $L\ge 0$ and consider a DGF $\Phi$.
Then, the global worst-case convergence time \eqref{eq:defTb} of system \eqref{eq:diffe} satisfies
  \begin{equation}
\Tb\Bigl(\Phi, \D_{\alpha, \beta} \ka, \alpha^2 L\Bigr) = \frac{1}{\alpha \beta} \Tb\Bigl(\Phi, \ka, L\Bigr).
  \end{equation}
  with $\D_{\alpha,\beta} = \operatorname{diag}(\alpha, \alpha^2, \beta)$ for all $\alpha, \beta \in \RR_{> 0}$.
\end{prp}

Proposition~\ref{prop:tuning} shows how the global convergence time depends on specific scalings of the parameters $\ka$ and the Lipschitz constant $L$. 
The following result allows, in addition, to bound the worst-case convergence time corresponding to a positive $L$ in terms of that corresponding to $L=0$. 
Its proof is also given in Section~\ref{sec:global:scaling}.
\begin{prp}[Variation of Lipschitz Constant]
  \label{prop:LDL0}
  Let $L\ge 0$, $\ka = (k_1,k_2,k_3)\in\RR^3_{>0}$, consider an admissible DGF $\Phi$ with $D \ge 1$ as in Definition~\ref{def:dgf:admissible} and define
  \begin{equation}
      \label{eq:Lmax}
    \overline L = \begin{cases}
        \frac{k_2}{D} & \text{if }k_1^2 \ge 8 k_2 \\
        \frac{k_2}{D} \tanh \dfrac{\pi k_1}{2 \sqrt{8 k_2 - k_1^2}} & \text{if }k_1^2 < 8 k_2.
    \end{cases}
  \end{equation}
If $L< \overline L$, then
  \begin{equation}
      \label{eq:TbLTb0}
    \Tb(\Phi,\ka,L) \le \frac{\Tb(\Phi, \ka, 0)}{1 - L \overline L^{-1}}.
  \end{equation}
\end{prp}

\begin{remark}
    Note that $k_2 > L D$ is necessary for the conditions of Proposition~\ref{prop:LDL0} to be satisfied, while $k_2 > L$ is a necessary condition for the finite-time stability of \eqref{eq:diffe}.
    Thus, it is desirable to keep $D \ge 1$ as small as possible when designing or choosing an admissible DGF $\Phi$.
\end{remark}

In order to simplify tuning, the following notion of a normalized parameter triple is introduced.
\begin{defin}[Normalized Parameter Triple]
    A parameter triple $\ka = (k_1, k_2, k_3) \in \RR_{>0}^3$ is called normalized with respect to an admissible DGF $\Phi$, if there exists a $D$ satisfying Definition~\ref{def:dgf:admissible}, i.e.,
    \begin{equation}
        D \ge \sup_{x \in \RR} \frac{\abs{\Phi''(x)}}{2 \abs{\Phi'(x)}^3},
    \end{equation}
    such that $\overline L \ge 1$ in \eqref{eq:Lmax}.
\end{defin}
Since $\overline L$ can be interpreted as an upper bound for the Lipschitz constant $L$, a normalized parameter triple ensures that fixed-time convergence can be guaranteed for all $L \le 1$.

Proposition~\ref{prop:LDL0} shows that it is sufficient to consider the case $L = 0$ when studying upper bounds for the global convergence time.
As the final prerequisite for the proposed tuning procedure, the following proposition gives such a bound.
It is proven in Section~\ref{sec:global:upper}.
\begin{prp}[Global Convergence Time Bound]
    \label{prop:upperbound}
    Let $\ka = (k_1,k_2,k_3) \in \RR_{>0}^3$ and consider an admissible DGF $\Phi$ with $B, C \in \RR_{> 0}$ as in Definition~\ref{def:dgf:admissible}.
    If $k_1^2 \ge 8 k_2$, then $\Tb(\Phi, \ka, 0) \le \tilde T(\Phi, \ka)$ with
    \begin{equation}
        \label{eq:upperbound:Ttilde}
        \tilde T(\Phi, \ka) = 
        \begin{cases}
            \frac{\log \frac{k_1 + \sqrt{k_1^2 - 8 k_2}}{k_1 - \sqrt{k_1^2 - 8 k_2}}}{2  k_3 \sqrt{k_1^2 - 8 k_2}} \scriptstyle C + \frac{k_1^2 + 4 k_2}{2 k_1 k_2 k_3} B, & k_1^2 > 8 k_2, \\
            \noalign{\vspace{5pt}}
            \frac{C + 6 B}{k_1 k_3}, & k_1^2 = 8 k_2.
        \end{cases}
    \end{equation}
\end{prp}

By combining Propositions~\ref{prop:tuning} and~\ref{prop:LDL0}, the scalar parameters $k_1, k_2, k_3$ may be computed from a single, arbitrary convergence time bound obtained using Proposition~\ref{prop:upperbound}.
This is stated as the following theorem.
\begin{theorem}[Tuning]
    \label{th:main}
    Let a Lipschitz constant $L\ge 0$ and a desired convergence time $T>0$ be given.
    Consider an admissible DGF $\Phi$, let $\tilde\ka = (\tilde k_1, \tilde k_2, \tilde k_3) \in \RR_{>0}^3$ be a normalized parameter triple with respect to $\Phi$, and suppose that
\begin{equation}
            \label{eq:main:bound}
            \Tb(\Phi, \tilde\ka, 0) \le \tilde T
        \end{equation}
holds for some $\tilde T > 0$.
    Define $\ka = (k_1, k_2, k_3)$ via
    \begin{align}
        \label{eq:tuning}
        k_1 &= \tilde k_1 \sqrt{\gamma}, &
        k_2 &= \tilde k_2 \gamma,&
        k_3 &= \frac{\tilde k_3 \sqrt{\gamma}}{\gamma - L} \cdot \frac{\tilde T}{T}
    \end{align}
    for some $\gamma > L$.
    Then, the global convergence time of the differentiator \eqref{eq:diff} satisfies the bound
    \begin{align}
      \Tb(\Phi,\ka,L) \le T.
    \end{align}
\end{theorem}
\begin{remark}
    Note that $\gamma$ acts as a tradeoff parameter that determines the relative magnitude of the parameters $k_1, k_2, k_3$ without influencing the convergence time.
    With increasing $\gamma$, the parameter $k_3$ decreases, while the parameters $k_1, k_2$, which according to Theorem~\ref{th:asymptotic} determine the behavior of system \eqref{eq:diffe} close to the origin, increase.
\end{remark}
\begin{remark}
    Table~\ref{tab:main:tuples} lists normalized parameter tuples for two admissible DGFs, along with the bound $\tilde T$ in \eqref{eq:main:bound} that may be used for differentiator tuning with this theorem.
The table's contents are derived in Section~\ref{sec:design-example}.
\end{remark}
\begin{pf}
    According to Proposition~\ref{prop:tuning}
    \begin{align}
        \Tb(\Phi, \ka, L) = \frac{\gamma - L}{\gamma} \frac{T}{\tilde T} \Tb(\Phi, \tilde\ka, \gamma^{-1} L).
    \end{align}
    Furthermore, since $\tilde{\ka}$ is normalized, Proposition~\ref{prop:LDL0} may be applied with $\overline L \ge 1$ to obtain
    \begin{equation}
        \Tb(\Phi, \tilde \ka, \gamma^{-1} L) \le \frac{1}{1 - \gamma^{-1} L \overline L^{-1}} \Tb(\Phi, \tilde \ka, 0) \le \frac{\gamma}{\gamma - L} \tilde T.
    \end{equation}
    Combining the two relations completes the proof.
    \QED
\end{pf}

\begin{table}[tb]
    \centering
\begin{tabular}{|l||c|c|c|c|c|}
        \hline
        \diagbox{$\Phi(x)$}{$\tilde k_1$} & $\sqrt{8}$ & $5$ & $10$ & $15$ & $20$ \\ \hline\hline
        $\spowf{x}{1}{2} + \spowf{x}{3}{2}$ & $6.9$ & $9.3$ & $16.5$ & $24.1$ & $31.9$ \\ \hline
    $\sqrt{\ee^{\abs{x}} - 1} \spow{x}{0}$ &  $7.1$ & $9.4$ & $16.6$ & $24.2$ & $31.9$ \\\hline
    \end{tabular}
    \caption{Gain $\tilde k_1$ and corresponding bound $\tilde T$, computed using \eqref{eq:main:bound}, of normalized parameter triples with $\tilde k_2 = \tilde k_3 = 1$ for two admissible DGFs, to be used for differentiator tuning with Theorem~\ref{th:main}}
    \label{tab:main:tuples}
\end{table}

\subsection{Tightness of Predefined Convergence Time}

By using the lower bound for $\Tb(\Phi, \ka, L)$ from Proposition~\ref{prop:motivation}, the conservatism of the predefined convergence time may be bounded from above, provided that the value of the improper integral in Definition~\ref{def:dgf:admissible} is known exactly:
\begin{theorem}[Tightness of Assigned Bound]
    \label{prop:worstcase}
    Let $T > 0$ and $\gamma > L \ge 0$, consider an admissible DGF $\Phi$, and suppose that $B$ is such that equality holds in item \eqref{it:Phi:integral} of Definition~\ref{def:dgf:admissible}.
    Suppose that $\tilde k_1$, $\tilde k_2$, $\tilde k_3$, and $\tilde T$ satisfy the conditions of Theorem~\ref{th:main} and that $\tilde k_1^2 \ge 8 \tilde k_2$.
    If parameters $k_1, k_2, k_3$ are selected using \eqref{eq:tuning}, then the ratio of predefined and actual global convergence time is bounded from above by
    \begin{equation}
        \label{eq:bound:conservativeness}
        \frac{T}{\Tb(\Phi, \ka, L)} \le
        \frac{\Bigl(\tilde k_1 - \sqrt{\tilde k_1^2 - 8 \tilde k_2}\Bigr) \tilde k_3 \tilde T}{2 B} \cdot \frac{\gamma}{\gamma - L}.
    \end{equation}
    Moreover, if $\tilde T = \tilde T(\Phi, \tilde \ka)$ from \eqref{eq:upperbound:Ttilde}, then this upper bound tends to~$\frac{\gamma}{\gamma - L}$ for $\tilde k_1 \to \infty$.
\end{theorem}
\begin{remark}
As a consequence, provided that $B$ is tight and \eqref{eq:upperbound:Ttilde} is used, the assigned convergence time bound can be made arbitrarily tight by increasing $\tilde k_1$ and $\gamma$.
\end{remark}
\begin{pf}
    Abbreviate $\tilde h = \sqrt{\tilde k_1^2 - 8 \tilde k_2}$ for convenience.
    Rewriting $T$ in \eqref{eq:tuning}, one has
    \begin{equation}
        T = \frac{\tilde k_3 \tilde T \sqrt{\gamma}}{(\gamma - L) k_3}. \end{equation}
    Definition~\ref{def:dgf:admissible}, item \eqref{it:Phi:integral} and Proposition~\ref{prop:motivation} furthermore yield
    \begin{equation}
        \Tb(\Phi, \ka, L) \ge 
\frac{2B}{\Bigl( k_1 - \sqrt{k_1^2 - 8 k_2} \Bigr) k_3} = \frac{2B}{\Bigl( \tilde k_1 - \tilde h \Bigr) \sqrt{\gamma} k_3}.
    \end{equation}
Dividing the two expressions yields the bound \eqref{eq:bound:conservativeness}.
Now, substitute $\tilde q_1 = \tilde k_2 \tilde k_1^{-2}$ and $\tilde q_2 = (\tilde k_1 + \tilde h) (\tilde k_1 - \tilde h)^{-1}$ in the multipliers of $B$ and $C$ in the expression $(\tilde k_1 - \tilde h) \tilde k_3 \tilde T(\Phi, \tilde \ka)$, and apply L'H\^opital's rule for $\tilde q_1 \to 0$ and $\tilde q_2 \to \infty$ to show that they tend to two and zero, respectively, as $\tilde k_1 \to \infty$.
\QED
\end{pf}

\section{Design Examples}
\label{sec:design-example}

\subsection{Uniform Robust Exact Differentiator}

As an application of the proposed tuning procedure, consider the uniform robust exact differentiator \citep{crumor_tac11}.
It is generated by the DGF $\Phi(x) = \spowf{x}{1}{2}  + \spowf{x}{3}{2}$ and is obtained from \eqref{eq:diff} and \eqref{eq:def:nu1nu2} as
\begin{subequations}
    \label{eq:diff:exmp}
    \begin{align}
        \dot y_1 &= k_1 (\spowf{f-y_1}{1}{2} + k_3^2 \spowf{f-y_1}{3}{2}) + y_2 \\
      \dot y_2 &= k_2 (\spow{f-y_1}{0} + 4 k_3^2 (f-y_1) + 3 k_3^4 \spow{f-y_1}{2}).
    \end{align}
\end{subequations}This DGF is admissible with
\begin{subequations}
    \label{eq:ured:BCD}
    \begin{align}
        \label{eq:ured:B}
        B &= \int_{0}^{\infty} \frac{1}{\Phi(x)} \diffd x = \int_{0}^{\infty} \frac{\diffd x}{x^{\frac{1}{2}} + x^{\frac{3}{2}}}
= \pi, \\
        C &= \sup_{x \in \RR} \frac{1}{\Phi'(x)} = \sup_{x > 0} \frac{1}{\frac{1}{2} x^{-\frac{1}{2}} + \frac{3}{2} x^{\frac{1}{2}}} = \frac{1}{\sqrt{3}}, \\
        D &= \sup_{x \in \RR} \frac{\abs{\Phi''(x)}}{2 \abs{\Phi'(x)}^3} 
= \sup_{x > 0} \frac{\abs{1 - 3 x}}{\abs{1 + 3 x}^3} = 1.
\end{align}
\end{subequations}

Using \eqref{eq:ured:BCD} and Proposition~\ref{prop:upperbound}, one obtains the normalized parameter triples and associated convergence time bounds in the first row of Table~\ref{tab:main:tuples}.
Choosing for example the first of these entries, a predefined convergence time $T$ may be assigned to the uniform robust exact differentiator by selecting $k_1, k_2, k_3$ according to Theorem~\ref{th:main} as
\begin{equation}
    \label{eq:ured:tuning:general}
    k_1 = \sqrt{8 \gamma}, \qquad
    k_2 = \gamma, \qquad
    k_3 = \frac{6.9 \sqrt{\gamma}}{(\gamma - L) T}.
\end{equation}

\begin{figure}[tbp]
    \centering
    \includegraphics{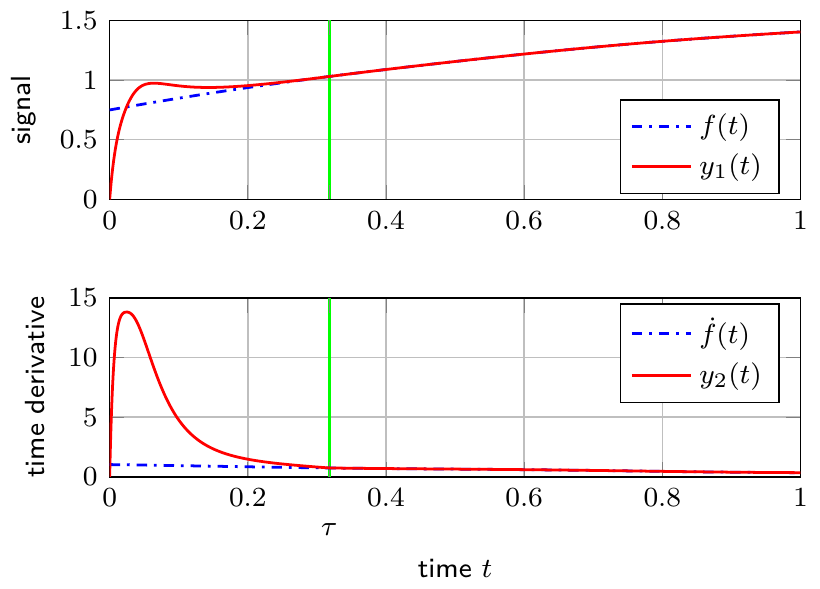}
    \caption{Differentiation of $f(t) = 0.75 \cos(t) + 0.0025 \sin(10 t) + t$ with $L = 1$ and desired convergence time bound $T = 1$ by using the proposed differentiator with parameters $k_1 = 6$, $k_2 = 4.5$, $k_3 \approx 4.182$, initial values $y_1(0) = y_2(0) = 0$, and DGF $\Phi(x) = \spow{x}{1/2} + \spow{x}{3/2}$, along with the actual convergence time $\tau \approx 0.32$, obtained using forward Euler discretization with discretization step size $T\sbrm{s} = 10^{-4}$.}
    \label{fig:diff_sim}
\end{figure}

Fig.~\ref{fig:diff_sim} shows simulation results obtained by applying \eqref{eq:diff:exmp} with initial values $y_1(0) = y_2(0) = 0$ to the signal 
\begin{equation}
    f(t) = 0.75 \cos(t) + 0.0025 \sin(10 t) + t
\end{equation}
with $L = 1$ and desired convergence time bound $T = 1$.
The parameters are chosen according to \eqref{eq:ured:tuning:general} with $\gamma = 4.5 L$ as $k_1 = 6$, $k_2 = 4.5$, and $k_3 \approx 4.182$.
For simplicity, the simulation is performed using forward Euler discretization using a sufficiently small step size $T\sbrm{s} = 10^{-4}$.
Note, however, that such a simple discretization scheme in general does not achieve global stability, as shown by \cite{levant2013fixed}.
In practice, more advanced schemes as proposed by \cite{wetzlinger2019semi} (cf. also \cite{rudiger2021robust}), for example, should therefore be employed.

Since $B$ is computed in \eqref{eq:ured:B} by solving the integral exactly, Theorem~\ref{prop:worstcase} may be applied to show that the assigned bound exceeds the worst-case convergence time by no more than a factor of four;
this can also be seen in the simulation.

\begin{figure}[tbp]
    \centering
    \includegraphics{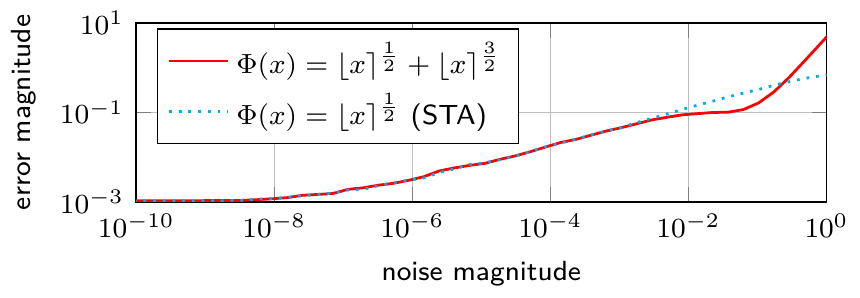}
    \caption{Comparison of steady-state differentiation error magnitude $\sup_{t \in [2,4]} \abs{x_2(t)}$ for the simulation shown in Fig.~\ref{fig:diff_sim} to the STA (i.e., the robust exact differentiator), with additional, uniformly bounded random measurement noise.}
    \label{fig:noise_sim}
\end{figure}

Fig.~\ref{fig:noise_sim} compares the steady-state differentiation error magnitude achieved using the proposed approach to the robust exact differentiator (i.e., the STA) using the same simulation settings with additional, uniformly bounded measurement noise.
As noise, a uniform random number independently sampled with step size $T\sbrm{s}$ is used.
For small noise, the behaviors coincide, as shown in Theorem~\ref{th:asymptotic}, and for vanishing noise they are determined by the discretization.
Only for very large noises, performance of the fixed-time differentiator eventually deteriorates due to the effectively larger gain that is necessary for achieving fixed-time convergence.

\subsection{Exponential Differentiator Generating Function}

As another example, consider the DGF
\begin{equation}
    \Phi(x) = \sqrt{\exp(\abs{x}) - 1} \spow{x}{0},
\end{equation}
which generates the differentiator
\begin{subequations}
    \label{eq:diff:exmp2}
    \begin{align}
        \dot y_1 &= \frac{k_1}{k_3} \sqrt{\exp(k_3^2 \abs{f - y_1}) - 1} \spow{f - y_1}{0} + y_2 \\
        \dot y_2 &= k_2 \exp(k_3^2 \abs{f - y_1}) \spow{f - y_1}{0}.
    \end{align}
\end{subequations}
For this DGF, (tight) admissibility constants are obtained as
\begin{subequations}
    \begin{align}
        \label{eq:exmp2:B}
        B &= \int_{0}^{\infty} \frac{1}{\Phi(x)} \diffd x = \int_{0}^{\infty} \frac{\diffd x}{\sqrt{\ee^{x}-1}}
= \pi, \\
        C &= \sup_{x \in \RR} \frac{1}{\Phi'(x)} = \sup_{x > 0} \frac{2 \sqrt{\ee^{x} - 1}}{\ee^{x}}  
= 1, \\
        D &= \sup_{x \in \RR} \frac{\abs{\Phi''(x)}}{2 \abs{\Phi'(x)}^3} = \sup_{x > 0} \frac{\abs{2 \ee^{x} - 3}}{\ee^{2 x}} = 1.
    \end{align}
\end{subequations}
Using these constants and Proposition~\ref{prop:upperbound}, the second row of Table~\ref{tab:main:tuples} is obtained.

\begin{figure}[tbp]
    \centering
    \includegraphics{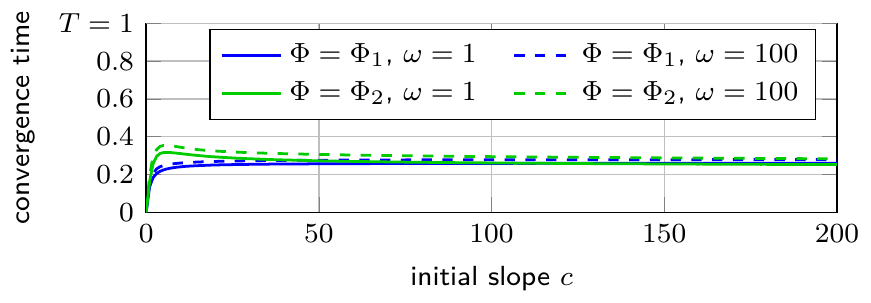}
    \caption{Simulated convergence times obtained by differentiating $f(t) = (\cos \omega t - 1)/\omega^2 + c t$ with settings as in Fig.~\ref{fig:diff_sim} and two different DGFs $\Phi_1(x) = \spowf{x}{1}{2} + \spowf{x}{3}{2}$ and $\Phi_2(x) = \sqrt{\ee^{\abs{x}} - 1} \spow{x}{0}$, both tuned according to Theorem~\ref{th:main} with convergence time bound $T = 1$ and $\gamma = 4.5$, as a function of the initial slope $c = \dot f(0)$.}
    \label{fig:convtime_comparison}
\end{figure}

Fig.~\ref{fig:convtime_comparison} depicts convergence times obtained from a simulation for both DGFs when differentiating functions $f(t)$ with different slope and sinusoidal frequency.
Simulation settings and parameters were selected as before, but with re-tuned parameter $k_3 \approx 4.303$ for the exponential DGF to maintain $T = 1$ as a global convergence time bound.
    The variation of the initial slope corresponds to a variation of the error system's initial condition $x_2(0) = \dot f(0)$, while\footnote{Variations of $x_1(0)$ are not considered here, because $x_1(0) = 0$ can always be achieved in practice by setting $y_1(0) = f(0)$.} $x_1(0) = 0$.
One can see that the convergence time remains bounded by $T$ for all considered functions.

\section{Convergence Time Function}
\label{sec:conv-time-func}

This section studies the convergence time of system \eqref{eq:diffe} as a function of the initial state.
First, this function is derived for the unperturbed case and for any DGF (even non-admissible ones) in the form of a convergent improper integral.
For admissible DGFs, a convergence time bound for the perturbed case is then shown.

Throughout this section as well as the next section, the following abbreviations are used.
Given parameters $\ka = (k_1,k_2,k_3)$, define the matrix $\A \in \RR^{2 \times 2}$, vectors $\e_1, \e_2 \in \RR^2$, and functions $\g : \RR^2 \to \RR^2$ and $\Psi : \RR \to \RR$ as
\begin{subequations}
    \label{eq:AgPsi}
    \begin{align}
        \label{eq:A}
        \A &= \begin{bmatrix}
            -\frac{k_1}{2} & \frac{1}{2} \\
            -k_2 & \ 0
        \end{bmatrix}, &
        \e_1 &= \begin{bmatrix}
            1 \\ 0
        \end{bmatrix}, &
        \e_2 &= \begin{bmatrix}
            0 \\ 1
        \end{bmatrix},\\
        \label{eq:gPsi}
        \g(\x) &= \begin{bmatrix}
            \Phi_{k_3}(x_1) \\
            x_2
        \end{bmatrix}, & \Psi(&\lefteqn{z) = \Phi_{k_3}^{-1}(z),}
    \end{align}
\end{subequations}
wherein according to \eqref{eq:Phieps}, $\Phi_{\epsilon}^{-1}(z) = \epsilon^{-2} \Phi^{-1}(\epsilon z)$.

Some basic properties of the function $\Psi$ are first shown in the following technical lemma, which is proven in Section~\ref{sec:proofs}.
\begin{lemma}
    \label{lem:Psi:bounds}
    Consider a DGF $\Phi$ let $k_3 \in \RR_{> 0}$.
    Then, for every $\epsilon > 0$ there exist constants $C, D \in \RR_{> 0}$ such that $\Psi$ as defined in \eqref{eq:AgPsi} satisfies for all $z \in [-\epsilon, \epsilon]$ the inequalities
    \begin{equation}
        \label{eq:Psi:bounds}
        |\Psi'(z)| \le C k_3^{-1}, \qquad
        |\Psi''(z)| \le 2D.
    \end{equation}
    Moreover, if $\Phi$ is an admissible DGF, then \eqref{eq:Psi:bounds} holds for all $z$ with $C, D$ as in Definition~\ref{def:dgf:admissible}.
\end{lemma}

\subsection{Computation for the Unperturbed Case}
\label{sec:reach-time-unpert}

Consider system \eqref{eq:diffe} without perturbation, i.e., with $\ddot f = 0$.
It is first shown that the corresponding convergence time function may be represented as an improper integral, which is finite for all positive parameters $\ka$ and all DGFs $\Phi$ that satisfy Definition~\ref{def:dgf}.

Following the ideas presented in \cite{crumor_tac11}, system \eqref{eq:diffe} with $\ddot f = 0$ and $\z = \g(\x)$ may be written as
\begin{equation}
    \dot{\z} = \frac{2}{\Psi'(z_1)} \A \z.
\end{equation}
As pointed out in \cite{seehor_tac18}, a linear system $\deriv{\z}{\tau} = \A \z$ is obtained with respect to a new time variable $\tau$ satisfying $\Psi'(z_1) \diffd \tau = 2 \diffd t$ \citep[this time-scaling idea was employed for the STA for the first time in][]{moreno2011new}.
Therefore, one may intuitively expect to obtain the convergence time by integrating $\diffd t$ along the trajectories of this linear system.

The following technical result, which is proven in Section~\ref{sec:proofs}, suggests that this is indeed the case.
\begin{lemma}
    \label{lem:V:unperturbed}
    Let $\ka = (k_1,k_2,k_3) \in \RR_{>0}^3$, and consider a DGF $\Phi$.
    Then, the function $V : \RR^2 \to \RR_{\ge 0}$ given by
    \begin{equation}
        V(\x) = \int_{0}^{\infty} \frac{1}{2} \Psi'(\e_1\TT \ee^{\A \tau} \g(\x)) \, \diffd \tau
    \end{equation}
    is locally bounded, continuous, and positive definite.
    For $x_1 \ne 0$, it is furthermore differentiable and its time derivative $\dot V$ along the trajectories of system \eqref{eq:diffe} is given by 
    \begin{equation*}
        \dot V(t, \x) = -1 + \ddot f(t) \int_{0}^{\infty} \frac{1}{2} \Psi''(\e_1\TT \ee^{\A \tau} \g(\x)) \e_1\TT \ee^{\A \tau} \e_2 \, \diffd \tau.
    \end{equation*}
\end{lemma}

The fact that $\dot V$ is equal to minus one for $\ddot f = 0$ suggests that the function $V$ is the unperturbed system's convergence time function.
Since $V$ is not everywhere differentiable, however, the following technical lemma is required, which is obtained following ideas presented in \cite{seehor_tac18,haideb_auto19}, and is proven in Section~\ref{sec:proofs}.
\begin{lemma}
    \label{lem:Vcomparison}
    Let $\ka \in \RR_{>0}^3$, $L \in \RR_{\ge 0}$, $c \in \RR_{>0}$ and consider a DGF $\Phi$.
    Let $V : \RR^2 \to \RR$ be continuous and positive definite and suppose that the time derivative $\dot V$ along the trajectories of system \eqref{eq:diffe} is well-defined and bounded by $\dot V(t, \x) \le -c$ for all $t, \x$ with $x_1 \ne 0$.
    Then, the system's convergence time $T_L^{\Phi,\ka}$ is bounded by
    \begin{equation}
        T_L^{\Phi,\ka}(\x) \le \frac{V(\x)}{c}
    \end{equation}
    and equality holds if $\dot V(t, \x) = -c$ for all $t, \x$ with $x_1 \ne 0$.
\end{lemma}

Using this result, the following proposition may now be proven.
\begin{prp}[Unperturbed Convergence Time]
  \label{prop:T:unperturbed}
  Let $\ka = (k_1,k_2,k_3) \in \RR_{>0}^3$, $\x_0 \in \RR^2$ and consider a DGF $\Phi$.
  Then, the convergence time of system \eqref{eq:diffe} for $L = 0$ is finite and is given by
  \begin{equation}
      \label{eq:T:unperturbed}
      T_0^{\Phi,\ka}(\x_0) = \int_{0}^{\infty} \frac{1}{2} \Psi'(\e_1\TT \ee^{\A \tau} \g(\x_0)) \, \diffd \tau.
  \end{equation}
\end{prp}
\begin{pf}
    For $\ddot f = 0$, the continuous, positive definite function $V$ defined in Lemma~\ref{lem:V:unperturbed} satisfies $\dot V(t, \x) = -1$ for $x_1 \ne 0$.
    Lemma~\ref{lem:Vcomparison} thus yields $T_0^{\Phi,\ka}(\x_0) = V(\x_0)$.
    Since this function is locally bounded according to Lemma~\ref{lem:V:unperturbed}, one has $T_0^{\Phi,\ka}(\x_0) < \infty$.
    \QED
\end{pf}

\subsection{Bound for the Perturbed Case}

Focusing on admissible DGFs, an upper bound for the convergence time function in the perturbed case is now derived.
To that end, the following lemma will be used, which expresses the maximum Lipschitz constant $\overline L$ defined in Proposition~\ref{prop:LDL0} in the form of an improper integral.
\begin{lemma}[{\citet[Theorem 3]{seeber2020computing}}]
    \label{lem:Lmax}
    Let $k_1, k_2, D \in \RR_{>0}$.
    Then, $\overline L$ defined in \eqref{eq:Lmax} is given by
    \begin{equation}
        \overline L = \frac{1}{D \int_{0}^{\infty} |\e_1\TT \ee^{\A \tau} \e_2| \, \diffd \tau}.
    \end{equation}
\end{lemma}

Using the auxiliary lemmas, an upper bound for the convergence time function in the perturbed case is obtained.
\begin{prp}[Convergence Time Bound]
    \label{prop:TLT0}
    Let $L\ge 0$, $\ka = (k_1,k_2,k_3)\in\RR^3_{>0}$, consider an admissible DGF $\Phi$, and define $\overline L$ as in \eqref{eq:Lmax}.
    If $L < \overline L$, then
  \begin{equation}
      \label{eq:TLT0}
      T_L^{\Phi,\ka}(\x) \le \frac{T_0^{\Phi,\ka}(\x)}{1 - L \overline L^{-1}}
  \end{equation}
  holds for all $\x \in \RR^{2}$.
\end{prp}
\begin{pf}
    Let $V$ be the continuous, positive definite function $V$ defined in Lemma~\ref{lem:V:unperturbed}.
    For $x_1 \ne 0$, this function is differentiable and using Lemmas~\ref{lem:V:unperturbed}, \ref{lem:Psi:bounds}, and~\ref{lem:Lmax} one finds that its time derivative along the trajectories of system \eqref{eq:diffe} is bounded by
    \begin{align}
        \dot V &= -1 + \ddot f \int_{0}^{\infty} \frac{1}{2} \Psi''(\e_1\TT \ee^{\A \tau} \g(\x)) \e_1\TT \ee^{\A \tau} \e_2 \, \diffd \tau \nonumber \\
        &\le -1 + L D \int_{0}^{\infty} |\e_1\TT \ee^{\A \tau} \e_2| \, \diffd \tau = -1 + L \overline L^{-1}.
    \end{align}
    Lemma~\ref{lem:Vcomparison} then yields
    \begin{equation}
        T_L^{\Phi,\ka}(\x) \le \frac{V(\x)}{1 - L \overline L^{-1}}
    \end{equation}
    and using Proposition~\ref{prop:T:unperturbed} to see that $V(\x) = T_0^{\Phi,\ka}(\x)$ concludes the proof.
    \QED
\end{pf}

\section{Global Convergence Time}
\label{sec:global-conv-time}

This section investigates properties and bounds of the convergence time function's smallest upper bound: the global convergence time
\begin{equation}
    \label{eq:global:Tb}
    \Tb(\Phi, \ka, L) = \sup_{\x \in \RR^{2}} T_L^{\Phi, \ka}(\x).
\end{equation}
First, two scaling properties are shown.
Then, lower and upper bounds are derived.
As before, the quantities introduced in~\eqref{eq:AgPsi} are used throughout this section.

\subsection{Scaling Properties}
\label{sec:global:scaling}

In the following, the scaling properties in Propositions~\ref{prop:tuning} and~\ref{prop:LDL0} are shown.
The former utilizes a homogeneity-like scaling property of state, time, and parameters of system \eqref{eq:diffe}.
The latter is an immediate consequence of Proposition~\ref{prop:TLT0}.

\paragraph*{PROOF of Proposition~\ref{prop:tuning}}
Let $x_1, x_2 : \RR_{\ge 0} \to \RR$ be any solution of \eqref{eq:diffe} with $|\ddot f| \le L$ and denote their convergence time by $\tau$.
Consider functions $z_1, z_2 : \RR_{\ge 0} \to \RR$ given by
\begin{equation}
    z_1(t) = \beta^{-2} x_1(\alpha \beta t), \qquad
    z_2(t) = \alpha \beta^{-1} x_2(\alpha \beta t).
\end{equation}
Clearly, these converge to zero in time $\tilde\tau = \alpha^{-1}\beta^{-1} \tau$ and satisfy
\begin{align}
    \dot z_1 &= \alpha \beta^{-1} \dot x_1 = - \alpha k_1 \beta^{-1} \Phi_{k_3}(x_1) + \alpha \beta^{-1} x_2 \nonumber \\
    &= - \alpha k_1 \beta^{-1} \Phi_{k_3}(\beta^2 z_1) + z_2 \nonumber \\ 
    &= - \alpha k_1 \Phi_{\beta k_3}(z_1) + z_2 \\
    \dot z_2 &= \alpha^2 \dot x_2 = - 2 \alpha^2 k_2 \Phi_{k_3}(x_1) \Phi_{k_3}'(x_1) + \alpha^2 \ddot f \nonumber \\
    &= - 2 \alpha^2 k_2 \Phi_{k_3}(\beta^2 z_1) \Phi_{k_3}'(\beta^2 z_1) + \alpha^2 \ddot f \nonumber \\
    &= -2 \alpha^2 k_2 \Phi_{\beta k_3}(z_1) \Phi_{\beta k_3}'(z_1) + \alpha^2 \ddot f,
\end{align}
i.e., differential equations \eqref{eq:diffe} with parameters $\alpha k_1$, $\alpha^2 k_2$, $\beta k_3$ and Lipschitz constant $\alpha^2 L$.
\QED

\paragraph*{PROOF of Proposition~\ref{prop:LDL0}}
Taking the supremum with respect to $\x$ on each side of \eqref{eq:TLT0} yields
\begin{align}
    \Tb(\Phi, \ka, L) &= \sup_{\x \in \RR^2} T_L^{\Phi,\ka}(\x) \nonumber \\
    &\le \sup_{\x \in \RR^2} \frac{T_0^{\Phi,\ka}(\x)}{1 - L \overline L^{-1}} = \frac{\Tb(\Phi, \ka, 0)}{1 - L \overline L^{-1}},
\end{align}
i.e., relation \eqref{eq:TbLTb0}, which proves the proposition.
\QED

\subsection{Lower Bound}
\label{sec:global:lower}

Explicitly solving the integral in \eqref{eq:T:unperturbed} is not possible in general.
An interesting special case is obtained if $\g(\x)$ is an eigenvector of $\A$ with eigenvalue $\lambda$.
In this case, one has
\begin{equation}
    \e_1\TT \ee^{\A \tau} \g(\x) = c \ee^{\lambda \tau}
\end{equation}
with some $c \in \RR$.
As the following lemma shows, the integral in \eqref{eq:T:unperturbed} may then be simplified further.
Its proof is given in Section~\ref{sec:proofs}.

\begin{lemma}
    \label{lem:singleexponential}
    Let $k_3 \in \RR_{> 0}$, $\lambda \in \RR_{< 0}$, $c \in \RR$ and consider a DGF $\Phi$.
    Then,
    \begin{equation}
        \int_{0}^{\infty} \Psi'(c \ee^{\lambda \tau}) \, \diffd \tau = -\frac{1}{k_3 \lambda} \int_{0}^{\Phi^{-1}(k_3 c)} \frac{1}{\Phi(x)} \, \diffd x.
    \end{equation}
\end{lemma}

It should be highlighted that the use of this lemma does not require the knowledge of the DGF's inverse $\Phi^{-1}$.
Proposition~\ref{prop:motivation}, which provided the original motivation for the definition of an admissible DGF, may now be proven.

\paragraph*{PROOF of Proposition~\ref{prop:motivation}}
Since $k_1^2 \ge 8 k_2$, the matrix $\A$ has real eigenvalues and its largest eigenvalue is given by
\begin{equation}
    \lambda = \frac{-k_1 + \sqrt{k_1^2 - 8 k_2}}{4}.
\end{equation}
Denote by $\mathcal D$ the set of all $\x \in \RR^2$ such that $\g(\x)$ is an eigenvector of $\A$ with this eigenvalue.
Using Proposition~\ref{prop:T:unperturbed} and Lemma~\ref{lem:singleexponential}, one then has
\begin{align}
    \Tb(\Phi, \ka, L) &
\ge \sup_{\x \in \mathcal D} T_0^{\Phi, \ka}(\x) = \sup_{c \in \RR} \int_{0}^{\infty} \frac{1}{2} \Psi'(c \ee^{\lambda \tau}) \, \diffd \tau \nonumber \\
    &= \sup_{c \in \RR} -\frac{1}{2 k_3 \lambda} \int_{0}^{\Phi^{-1}(k_3 c)} \frac{1}{\Phi(x)} \, \diffd x \nonumber \\
&= \frac{2 \int_{0}^{\infty} \frac{1}{\Phi(x)} \, \diffd x}{\bigl( k_1 - \sqrt{k_1^2 - 8 k_2} \bigr) k_3}.
\end{align}This concludes the proof.\QED

\subsection{Upper Bound}
\label{sec:global:upper}

Upper bounds for the global convergence time are now studied.
According to the results in the previous sections, it is sufficient to consider such bounds for $L = 0$, since bounds for $L > 0$ can then be obtained using Proposition~\ref{prop:LDL0}.
Using Proposition~\ref{prop:T:unperturbed}, the expression to bound from above is given by
\begin{equation}
    \label{eq:Tb:supz}
    \Tb(\Phi, \ka, 0) = \sup_{\z \in \RR^{2}} \int_{0}^{\infty} \frac{1}{2} \Psi'(\e_1\TT \ee^{\A \tau} \z) \, \diffd \tau.
\end{equation}

As a first step towards finding the supremum, the following lemma, which is proven in Section~\ref{sec:proofs}, is given.
It allows to restrict the range of $\z$ to a compact subset of $\RR^2$ by extending the domain of integration.
In case the eigenvalues of $\A$ are real-valued and distinct, the integrand may furthermore be simplified.
\begin{lemma}
    \label{lem:convtime:alternateform}
    Consider an admissible DGF $\Phi$, define the compact set $\mathcal{D} = \{ \v \in \RR^{2} : \norm{\v} = 1 \}$, and let $\ka = (k_1,k_2,k_3) \in \RR_{>0}^3$.
    Then,
        \begin{equation}
            \label{eq:convtime:alternateform1}
        \Tb(\Phi, \ka, 0) = \sup_{\v \in \mathcal{D}} \int_{-\infty}^{\infty} \frac{1}{2} \Psi'(\e_1\TT \ee^{\A \tau} \v) \, \diffd \tau.
    \end{equation}
    Furthermore, if $k_1^2 > 8 k_2$, then
    \begin{equation}
    \label{eq:convtime:alternateform2}
        \Tb(\Phi, \ka, 0) = \sup_{a \in \RR} \int_{-\infty}^{\infty} \frac{1}{2} \Psi'(|a| \ee^{\lambda_1 \tau} + a \ee^{\lambda_2 \tau}) \, \diffd \tau
    \end{equation}
    holds with the eigenvalues $\lambda_1, \lambda_2 \in \RR_{< 0}$
    \begin{equation}
        \label{eq:A:eig}
        \lambda_1 = - \frac{k_1 - \sqrt{k_1^2 - 8 k_2}}{4}, \quad
        \lambda_2 = -\frac{k_1 + \sqrt{k_1^2 - 8 k_2}}{4}
    \end{equation}
    of the matrix $\A$.
\end{lemma}

In order to obtain global bounds, the following lemma is used, which is essentially an extension of Lemma~\ref{lem:singleexponential} and is proven in Section~\ref{sec:proofs}.
\begin{lemma}
    \label{lem:integralbound}
    Let $k_3 > 0$ and consider an admissible DGF $\Phi$ and a (possibly unbounded) interval $(a, b) \subseteq \RR$.
    Let furthermore $h : (a, b) \to \RR$ be a continuously differentiable function, which for some $\alpha \in \RR$ satisfies
    \begin{equation}
      \label{eq:hderbnd}
        h'(\tau) \ge \alpha h(\tau) > 0
    \end{equation}
    for all $\tau \in (a, b)$.
    Then, with $\Psi$ as defined in \eqref{eq:gPsi}, one has
    \begin{equation}
        \int_{a}^{b} \Psi'(h(\tau)) \, \diffd \tau \le \frac{1}{k_3 \alpha} \int_{\Phi^{-1}(k_3 h(a))}^{\Phi^{-1}(k_3 h(b))} \frac{1}{\Phi(x)} \, \diffd x.
    \end{equation}
\end{lemma}

Using these results, Proposition~\ref{prop:upperbound} may now be proven.
\paragraph*{PROOF of Proposition~\ref{prop:upperbound}:}
The second case $k_1^2 = 8 k_2$ in \eqref{eq:upperbound:Ttilde} is obtained as the limit of the first case as $k_1^2$ tends to $8 k_2$.
Hence, it is sufficient to consider the case $k_1^2 > 8 k_2$ in the following.
In this case, one has $\Tb(\Phi, \ka, 0) = \sup_{a \in \RR}  T_a$ according to Lemma~\ref{lem:convtime:alternateform} with
    \begin{equation}
        \label{eq:Ta}
        T_a = \int_{-\infty}^{\infty} \frac{1}{2} \Psi'(|a| \ee^{\lambda_1 \tau} + a \ee^{\lambda_2 \tau}) \, \diffd \tau
    \end{equation}
    and $\lambda_2 < \lambda_1$ as in \eqref{eq:A:eig}.
    Consider first the case $a \ge 0$.
    Since $\Psi'(-z) = \Psi'(z)$, one may apply Lemma~\ref{lem:integralbound} with $h(\tau) = -a \ee^{\lambda_1 \tau} - a \ee^{\lambda_2 \tau}$.
    This function satisfies $h(\tau) < 0$ and
    \begin{equation}
        h'(\tau) = \lambda_1 h(\tau) + a(\lambda_1 - \lambda_2) \ee^{\lambda_2 \tau} \ge \lambda_1 h(\tau) > 0
    \end{equation}
    for all $\tau \in \RR$, because $\lambda_2 < \lambda_1 < 0$.
    One thus obtains
    \begin{equation}
        \label{eq:Ta1}
        T_a \le \frac{\int_{-\infty}^{0} \frac{1}{\Phi(x)} \, \diffd x }{ 2 k_3 \lambda_1 } = - \frac{\int_{0}^{\infty} \frac{1}{\Phi(x)} \, \diffd x}{2 k_3 \lambda_1}  \le -\frac{B}{2 k_3 \lambda_1}.
\end{equation}
    For the case $a < 0$, let $b = -a > 0$ and consider the function
    \begin{equation}
        \tilde h_b(\tau) = |a| \ee^{\lambda_1 \tau} + a \ee^{\lambda_2 \tau} = b (\ee^{\lambda_1 \tau} - \ee^{\lambda_2 \tau}).
    \end{equation}
On the interval $I_1 = (-\infty, 0)$ this function is strictly increasing and satisfies in particular $\tilde h_b(\tau) < 0$ and
    \begin{equation}
        \tilde h'_b(\tau) = \lambda_2 \tilde h_b(\tau) + b (\lambda_1 - \lambda_2) \ee^{\lambda_1 \tau}  \ge - \frac{k_1}{4} \tilde h_b(\tau)
    \end{equation}
    for all $\tau \in I_1$.
    It furthermore has a single inflection point
    \begin{equation}
        \tau\sbrm{inf} = \frac{2}{\lambda_1 - \lambda_2} \log\frac{\lambda_2}{\lambda_1} = \frac{\log \frac{k_1 + \sqrt{k_1^2 - 8k_2}}{k_1 - \sqrt{k_1^2 - 8k_2}}}{\sqrt{k_1^2 - 8 k_2}},
    \end{equation}
    i.e., $\tilde h_b''(\tau\sbrm{inf}) = 0$.
    On the interval $I_3 = (\tau\sbrm{inf}, \infty)$, one has $\tilde h_b''(\tau) > 0$; thus
    \begin{equation}
        (\lambda_1 + \lambda_2) \tilde h_b'(\tau) = \lambda_1 \lambda_2 \tilde h_b(\tau) + \tilde h_b''(\tau) > \lambda_1 \lambda_2 \tilde h_b(\tau).
    \end{equation}
    and consequently
    \begin{equation}
        \tilde h_b'(\tau) \le \frac{\lambda_1 \lambda_2}{\lambda_1 + \lambda_2} \tilde h_b(\tau) = - \frac{k_2}{k_1} \tilde h_b(\tau) < 0
    \end{equation}
    holds for all $\tau \in I_3$.
    Therefore, Lemma~\ref{lem:integralbound} may be used with $h = \tilde h_b$ or $h = - \tilde h_b$ on the intervals $I_1$ or $I_3$, respectively, to bound the integral in \eqref{eq:Ta} from above.
    On the remaining interval $I_2 = (0, \tau\sbrm{inf})$, no suitable inequality may be found, because $\tilde h_b$ is positive while $\tilde h_b'$ changes sign.
    According to Lemma~\ref{lem:Psi:bounds}, the integrand is bounded by $\abs{\Psi'(z)} \le C k_3^{-1}$ on this finite interval, however, which together with the use of Lemma~\ref{lem:integralbound} on $I_1$ and $I_3$ yields the bound
    \begin{align}
        \label{eq:Ta2}
        T_a &\le -\frac{2}{k_3 k_1} \int_{-\infty}^{0} \frac{1}{\Phi(x)} \, \diffd x + \int_{0}^{\tau\sbrm{inf}} \frac{C}{2 k_3} \, \diffd \tau  \nonumber \\
        &\quad-\frac{k_1}{2 k_2 k_3} \int_{\Phi^{-1}(-k_3 \tilde h_b(\tau\sbrm{inf}))}^{0} \frac{1}{\Phi(x)} \, \diffd x \nonumber \\
        &\le \frac{\tau\sbrm{\inf}C}{2 k_3} + \Bigl( \frac{k_1}{2 k_2 k_3} + \frac{2}{k_3 k_1} \Bigr) \int_{0}^{\infty} \frac{1}{\Phi(x)} \, \diffd x \nonumber \\
        &\le \frac{\log \frac{k_1 + \sqrt{k_1^2 - 8k_2}}{k_1 - \sqrt{k_1^2 - 8k_2}}}{2 k_3 \sqrt{k_1^2 - 8 k_2} } C + \frac{k_1^2 + 4k_2}{2 k_3 k_1 k_2} B.
    \end{align}
    The proof is concluded by noting that
    \begin{equation}
        - \frac{1}{2 \lambda_1} = \frac{k_1 + \sqrt{k_1^2 - 8k_2}}{4 k_2} \le \frac{k_1}{2 k_2}
    \end{equation}
    holds and thus \eqref{eq:Ta1} implies \eqref{eq:Ta2}.
    \QED

\section{Conclusions and Outlook}
\label{sec:conclusions}

A class of fixed-time convergent differentiators was proposed for differentiating an arbitrary signal with Lipschitz continuous time derivative in a predefined finite time.
The differentiators are parameterized using a scalar differentiator generating function (DGF) and three scalar parameters.

Admissibility conditions for the DGF were given, and proper selection of the DGF was shown to yield existing differentiators, such as the uniform robust exact differentiator, as special cases.
The proposed tuning procedure allows to assign any predefined convergence time bound by computing the three scalar differentiator parameters using a simple tuning rule.
The assigned bound can furthermore be made arbitrarily tight by appropriate selection of two tradeoff parameters that appear in the tuning rule.

For functions with constant derivative, the differentiator's convergence time was computed in the form of an improper integral.
It was shown that maximizing such an integral over a compact set yields the actual global convergence time, and an upper bound for it was derived in analytic form.

Future research may study the discrete-time implementation of the differentiator and further investigate its properties under large scale measurement noises.
Furthermore, possibilities  may be explored for solving the obtained optimization problem in closed form or for extending the proposed differentiator to obtain higher-order derivatives.

\section{Proofs}
\label{sec:proofs}

\subsection{Proof of Lemma~\ref{lem:nu1nu2:limits}}
To see that the limits hold pointwise, first note that
\begin{align}
    \label{eq:Phi:sign}
    \lim_{\alpha \to 0} \abs{\nu_2(\alpha^2 x)} &= \lim_{x \to 0} \abs{2 \Phi(x) \Phi'(x)} = \lim_{x \to 0^+} \frac{2 \Phi(x)}{\frac{1}{\Phi'(x)}}  \nonumber \\
&\hspace{-2em}= \lim_{x \to 0^+} \frac{2 \Phi'(x)}{-\frac{\Phi''(x)}{\Phi'(x)^2}} =  \lim_{x \to 0^+} \frac{2 \Phi'(x)^3}{-\Phi''(x)} = 1,
\end{align}
where L'H\^opital's rule may be applied because, due to items \eqref{it:Phi:differentiable}, \eqref{it:Phi:increasing}, and \eqref{it:Phi:limit} of Definition~\ref{def:dgf}, $\Phi(x)$ and $\frac{1}{\Phi'(x)}$ are continuously differentiable and tend to zero at the origin.
Using this relation and applying L'H\^opital's rule again yields
\begin{equation}
    \label{eq:galpha:lim}
    \lim_{\alpha \to 0} \frac{\nu_1(\alpha^2 x)^2}{\alpha^2} = \lim_{\alpha \to 0} \frac{\nu_2(\alpha^2 x) 2 \alpha x}{2 \alpha} = \abs{x}, \\
\end{equation}
where the fact that $\nu_2$ is odd is also used.

Uniformity of the limit \eqref{eq:Phi:sign}  is trivial, because for all $c > 0$
\begin{equation}
    \lim_{\alpha \to 0} \sup_{x \in (0, c]} \abs{\nu_2(\alpha^2 x) - 1} = \lim_{\alpha \to 0} \sup_{y \in (0, \alpha^2 c]} \abs{\nu_2(y) - 1} = 0
\end{equation}
follows from pointwise convergence.
To see uniformity of the other limit, consider the derivative $g_{\alpha}'$ of the relevant function family $g_{\alpha}(x) := \alpha^{-2} \nu_1(\alpha^2 x)^2$,
\begin{equation}
    g_{\alpha}'(x) = \deriv{}{x} \frac{\nu_1(\alpha^2 x)^2}{\alpha^2} = \nu_2(\alpha^2 x).
\end{equation}
According to item \eqref{it:Phi:differentiable} of Definition~\ref{def:dgf}, $\nu_2$ is continuously differentiable on $\RR \setminus \{ 0 \}$.
Since $\nu_2$ also stays bounded near the origin due to \eqref{eq:Phi:sign}, $g_{\alpha}'(x)$ is uniformly bounded with respect to $x$ \emph{and} $\alpha$ on all compact subsets of $\RR \times \RR$.
For every $x$, $g_{\alpha}(x)$ is moreover uniformly bounded with respect to $\alpha$ due to continuity for $\alpha \in \RR \setminus \{ 0\}$ and \eqref{eq:galpha:lim}.
Hence, the function family $g_{\alpha}(x)$ is uniformly equicontinuous and bounded on compact sets, which implies uniform convergence of $g_{\alpha}(x)$ to its pointwise limit.~\QED

\subsection{Proof of Lemma~\ref{lem:Psi:bounds}}

By differentiating $\Psi = \Phi_{k_3}^{-1}$ twice, one obtains
\begin{subequations}
    \begin{align}
(\Phi_{k_3}^{-1})' &= \frac{1}{\Phi_{k_3}' \circ \Phi_{k_3}^{-1}} &
(\Phi_{k_3}^{-1})'' &= -\frac{\Phi_{k_3}'' \circ \Phi_{k_3}^{-1}}{(\Phi_{k_3}' \circ \Phi_{k_3}^{-1})^3}
    \end{align}
\end{subequations}
Due to items \eqref{it:Phi:differentiable} and \eqref{it:Phi:increasing} of Definition~\ref{def:dgf}, these functions are continuous on $\RR \setminus \{ 0 \}$.
Define $\Psi'(0) = 0$ and $\Psi''(0) = 2$; then, due to items \eqref{it:Phi:limit} and \eqref{it:Phi:discont} of Definition~\ref{def:dgf}, respectively, it follows that $\Psi'$ and $\Psi''$ are continuous at 0 and hence on $\RR$.
In consequence, $\Psi'$ and $\Psi''$ are both uniformly bounded on every compact subset of $\RR$.

In case $\Phi$ is an admissible DGF, items \eqref{it:Phi:minimum} and \eqref{it:Phi:growth} of Definition~\ref{def:dgf:admissible} moreover yield
\begin{subequations}
    \begin{align}
        \abs{(\Phi_{k_3}^{-1})'(z)} &= \frac{1}{k_3 \Phi'(k_3^2 \Phi_{k_3}^{-1}(z))} \le \frac{C}{k_3} \\
\abs{(\Phi_{k_3}^{-1})''(z)} &= \frac{k_3^3 \abs{\Phi_{k_3}''(k_3^2 \Phi_{k_3}^{-1}(z))}}{|k_3 \Phi'(k_3^2 \Phi_{k_3}^{-1}(z))|^3} \le 2 D,
    \end{align}
\end{subequations}
since $\Phi_{k_3}'(x) = k_3 \Phi'(k_3^2 x)$ and $\Phi_{k_3}''(x) = k_3^3 \Phi''(k_3^2 x)$, proving validity of the bounds \eqref{eq:Psi:bounds} for all $z \in \RR$.
\QED

\subsection{Proof of Lemma~\ref{lem:V:unperturbed}}

Consider the function
$
    h(\x,\tau) = \e_1\TT \ee^{\A \tau} \g(\x)
    $.
The positive definiteness of $V$ follows from the facts that $\Psi'$ is positive definite and that $h(\x,\tau)$ cannot be zero for all $\tau \ge 0$ unless $\x = 0$. 
To show that $V(\x)$ is well-defined and locally bounded, consider any $\x \in \RR^2$. 
Since $\A$ is a Hurwitz matrix, the function $h(\x,\cdot)$ is uniformly bounded and converges to zero.
By item~(\ref{it:Phi:discont}) of Definition~\ref{def:dgf}
\begin{align}
    \lim_{z \to 0} \frac{2 \abs{z}}{\Psi'(z)} &= \lim_{z \to 0} 2 \abs{z \Phi_{k_3}'(\Phi_{k_3}^{-1}(z))}  \nonumber\\
    &= \lim_{x \to 0} 2 \abs{\Phi_{k_3}(x) \Phi_{k_3}'(x)} = 1 
\end{align}
holds, and therefore $\Psi'(z) \le (2+\varepsilon) \abs{z}$ holds for $\varepsilon > 0$, for sufficiently small values of $z$.
In particular, let $\delta_0 > 0$ be such that 
\begin{align}
  \label{eq:derPsiineq}
  0 \le \Psi'(z) \le 3 \abs{z},\quad \forall |z| \le \delta_0.
\end{align}
Since the integral $\int_0^{\infty} \abs{h(\x,\tau)} \diffd \tau$ converges, $V(\x)$ is finite.

Continuity of $V$ in $\RR^2$ will next be established by establishing uniform continuity in every closed ball $B_r = \{ \x \in \RR^2 : \|\x \| \le r \}$.
Note that \eqref{eq:derPsiineq} actually shows that $\lim_{z\to 0} \Psi'(z) = 0 = \Psi'(0)$, so that $\Psi'$ is continuous at 0.
In addition, from Definition~\ref{def:dgf} and (\ref{eq:gPsi}), it then follows that $\Psi'$ is continuous everywhere.
Since $\A$ is Hurwitz, the function $h$ has the following property: there exists $\lambda > 0$ such that for all $r\ge 0$, there exists $M = M(r) \ge 0$ so that
\begin{align}
  \label{eq:proph1}
  |h(\x,\tau)| \le M e^{-\lambda\tau} \quad \forall \x\in B_r, \forall \tau \ge 0.
\end{align}
Let $r > 0$, consider $M=M(r)$ as before, and let $\varepsilon > 0$.
Note that $h(\x,\tau) \in B_M$ for all $\x\in B_r$ and $\tau\ge 0$.
Define
\begin{align*}
    \varepsilon_2 = \min\left\{\delta_0, \frac{\varepsilon\lambda}{12} \right\}, \,
    T = \max\left\{ \frac{1}{\lambda} \log \frac{M}{\varepsilon_2}, 0 \right\},  \,
  \varepsilon_1 = \frac{\varepsilon}{2T}.
\end{align*}
From these definitions and (\ref{eq:proph1}), it follows that
\begin{align}
  |h(\x,\tau)| \le \varepsilon_2 e^{-\lambda (\tau-T)} \quad \forall \x\in B_r, \forall \tau \ge T.
\end{align}
Since $\Psi'$ is continuous, it is uniformly continuous in the closed ball $B_M$.
Then, there exists $\delta_1 = \delta_1(\varepsilon_1) > 0$ such that for all $\z_1,\z_2 \in B_M$, 
\begin{align*}
  |\Psi'(\z_1) - \Psi'(\z_2)| < \varepsilon_1\quad
  \text{if }|\z_1 - \z_2| < \delta_1.
\end{align*}
From the continuity of $h$, there exists $\delta = \delta(\delta_1,T) > 0$ such that 
$
  |h(\x_1,\tau) - h(\x_2,\tau)| < \delta_1
  $
for all $\x_1,\x_2 \in B_r$ and $0\le \tau \le T$ with $|\x_1 - \x_2| < \delta$.
Then, for every $\x_1,\x_2 \in B_r$ satisfying $|\x_1 - \x_2| < \delta$,
\begin{align*}
    \lefteqn{|V(\x_1) - V(\x_2)|}\hspace{1cm}\\ 
  &=\left|\int_0^\infty \Psi'(h(\x_1,\tau)) - \Psi'(h(\x_2,\tau)) d\tau\right|\\
  &\le \int_0^\infty \big|\Psi'(h(\x_1,\tau)) - \Psi'(h(\x_2,\tau))\big| d\tau\\
  &\le \int_0^T \big|\Psi'(h(\x_1,\tau)) - \Psi'(h(\x_2,\tau))\big| d\tau + \\ 
  &\phantom{\le}\int_T^\infty \big|\Psi'(h(\x_1,\tau))\big| + \big|\Psi'(h(\x_2,\tau))\big| d\tau\\
  &< T \varepsilon_1 + \int_T^\infty 3  |h(\x_1,\tau)| + 3|h(\x_2,\tau)| d\tau\\ \displaybreak[0]
  &\le \varepsilon/2 + 6 \varepsilon_2 \int_T^\infty e^{-\lambda (\tau-T)} d\tau\\
  &\le \frac{\varepsilon}{2} + 6 \frac{\varepsilon_2}{\lambda} \le \frac{\varepsilon}{2} + \frac{\varepsilon}{2} = \varepsilon
\end{align*}
This shows that $V$ is uniformly continuous in $B_r$.

To show differentiability for $x_1 \ne 0$, note that $\Psi''$ is uniformly bounded on any compact subset of $\RR$ according to Lemma~\ref{lem:Psi:bounds} for any DGF $\Phi$.
Since, for any given $\x \in \RR^2$, $h(\x, \tau)$ stays in such a compact subset, one obtains
\begin{equation}
    \label{eq:V:partial}
    \pderiv{V}{\x} = \int_{0}^{\infty} \frac{1}{2} \Psi''(\e_1\TT \ee^{\A \tau} \g(\x)) \e_1\TT \ee^{\A \tau} \pderiv{g}{\x} \, \diffd \tau,
\end{equation}
where Lebesgue's dominated convergence theorem may be used to show that differentiation and improper integration may be interchanged because $\Psi''(h(\x,\tau))$ is uniformly bounded with respect to $\tau$.
Using this relation to compute the time derivative $\dot V$ of $V$ along the trajectories of \eqref{eq:diffe} for $\ddot f = 0$ yields
\begin{align}
    \dot V(t, \x) &= \int_{0}^{\infty} \frac{\Psi''(\e_1\TT \ee^{\A \tau} \g(\x))}{\Psi'(\e_1\TT \g(\x))} \e_1\TT \ee^{\A \tau} \A \g(\x) \, \diffd \tau \nonumber \\
    &= \int_{0}^{\infty} \frac{\deriv{}{\tau} \Psi'(\e_1\TT \ee^{\A \tau} \g(\x))}{\Psi'(\e_1\TT \g(\x))} \, \diffd \tau  = -1.
\end{align}
Using this result, the claim follows by noting that $\dot V$ depends affinely on $\ddot f$ and by using \eqref{eq:diffe} and \eqref{eq:V:partial} to compute it.
\QED

\subsection{Proof of Lemma~\ref{lem:Vcomparison}}

Consider any $\x_0 \in \RR^2$, denote by $\tau := T_L^{\Phi,\ka}(\x_0)$ the corresponding convergence time.
Assume to the contrary that $\tau > c^{-1} V(\x_0)$.
Then, there exists $\epsilon > 0$ such that also $\tau - \epsilon > c^{-1} V(\x_0)$.
Let $\x(t)$ be a trajectory of the system that satisfies $\x(t) \ne 0$ for all $t \le \tau - \epsilon$ and consider this trajectory on the interval $\mathcal{I} = [0, \tau-\epsilon]$.
Since $x_1(t) = 0$ implies $\dot x_1(t) = x_2(t) \ne 0$ for all $t \in \mathcal{I}$, there is only a finite number of zero crossings of $x_1$ on the interval $\mathcal{I}$ and $\dot V(t, \x(t))$ exists for all but finitely many $t \in \mathcal{I}$.
It is therefore Henstock-Kurzweil integrable \citep[Theorem 5.7.7]{bogachev2007measure} and
\begin{equation*}
    V(\x(t)) = V(\x_0) + \int_{0}^{t} \dot V(\x(\tau)) \, \diffd \tau
\end{equation*}
for all $t \in \mathcal{I}$.
Since $\dot{V}$ is furthermore bounded from above, it is also Lebesgue integrable \citep[Corollary 5.7.11]{bogachev2007measure} and the integrals' values coincide \citep[cf.][Theorem 5.7.14]{bogachev2007measure}.
Therefore, one has
\begin{equation*}
    V(\x(\tau- \epsilon)) \le V(\x_0) - \int_{0}^{\tau-\epsilon} c \, \diffd \tau = V(\x_0) - (\tau - \epsilon) c < 0
\end{equation*}
which contradicts the fact that $V$ is positive definite.
Therefore, $\tau \le c^{-1} V(\x_0)$.
\QED

\subsection{Proof of Lemma~\ref{lem:singleexponential}}

Consider the substitution $y = \Psi(c \ee^{\lambda \tau})$ with
\begin{equation}
    \diffd y = c \lambda \ee^{\lambda \tau} \Psi'(c \ee^{\lambda \tau}) \, \diffd \tau = \lambda \Psi^{-1}(y) \Psi'(c \ee^{\lambda \tau}) \, \diffd \tau
\end{equation}
Using $\Psi = \Phi_{k_3}^{-1}$, one thus has
\begin{align*}
    \int_{0}^{\infty} \Psi'(c \ee^{\lambda \tau}) \, \diffd \tau &= \int_{\Psi(c)}^{\Psi(0)} \frac{1}{\lambda \Psi^{-1}(y)} \, \diffd y \nonumber \\
    &= - \frac{1}{\lambda} \int_{0}^{\Phi_{k_3}^{-1}(c)} \frac{1}{k_3^{-1} \Phi(k_3^2 y)} \, \diffd y.
\end{align*}
Since $\Phi_{k_3}^{-1}(c) = k_3^{-2} \Phi^{-1}(k_3 c)$, the substitution $x = k_3^2 y$ yields the claimed result.
\QED
\subsection{Proof of Lemma~\ref{lem:convtime:alternateform}}

To show \eqref{eq:convtime:alternateform1}, note that $\A$ is Hurwitz and, thus, for every $\z \in \RR^2 \setminus \{ \bm{0} \}$ there exists $\sigma \in \RR$ (depending on $\z$) such that $\v = \ee^{-\A \sigma} \z$ satisfies $\norm{\v} = 1$, i.e., $\v \in \mathcal{D}$.
Furthermore, the function mapping $\z$ to $\sigma$ is surjective.
Hence, \eqref{eq:Tb:supz} may be rewritten as
\begin{align}
    \label{eq:alternateform:Tb}
    \Tb(\Phi, \ka, 0) &= \sup_{\sigma \in \RR} \sup_{\v \in \mathcal{D}} \int_{0}^{\infty} \frac{1}{2} \Psi'(\e_1\TT \ee^{\A (\tau + \sigma)} \v) \, \diffd \tau \nonumber \\
    &= \sup_{\sigma \in \RR} \sup_{\v \in \mathcal{D}} \int_{\sigma}^{\infty} \frac{1}{2} \Psi'(\e_1\TT \ee^{\A \tau} \v) \, \diffd \tau \nonumber \\
    &= \sup_{\v \in \mathcal{D}} \int_{-\infty}^{\infty} \frac{1}{2} \Psi'(\e_1\TT \ee^{\A \tau} \v) \, \diffd \tau.
\end{align}

To show \eqref{eq:convtime:alternateform2}, suppose that $k_1^2 > 8 k_2$.
Then, the eigenvalues $\lambda_1, \lambda_2$ of $\A$ are real-valued and distinct.
Consequently, for every $\v \in \mathcal{D}$ there exist $a, \kappa \in \RR$ and $s \in \{ -1, 1\}$ such that the integrand in \eqref{eq:alternateform:Tb} has the form
\begin{equation}
    \e_1\TT \ee^{\A \tau} \v = s a \ee^{\lambda_1(\tau - \kappa)} + a \ee^{\lambda_2(\tau - \kappa)},
\end{equation}
and the function mapping $\v$ to $a$ is surjective.
Considering, due to the symmetry of $\Psi'$, only the cases where the expression $s a$ is non-negative, i.e., where $s a = \abs{a}$, one has
\begin{align}
\Tb(\Phi, \ka, 0) &= \sup_{a \in \RR} \int_{-\infty}^{\infty} \frac{1}{2} \Psi'(\abs{a} \ee^{\lambda_1(\tau - \kappa)} + a \ee^{\lambda_2(\tau - \kappa)}) \, \diffd \tau \nonumber \\
    &= \sup_{a \in \RR} \int_{-\infty}^{\infty} \frac{1}{2} \Psi'(\abs{a} \ee^{\lambda_1 \tau} + a \ee^{\lambda_2 \tau}) \, \diffd \tau,
\end{align}
which concludes the proof.

\subsection{Proof of Lemma~\ref{lem:integralbound}}

From~(\ref{eq:hderbnd}), it follows that $h$ is strictly increasing and hence invertible in the interval $(a,b)$. Consider the substitution $x = k_3^2\Psi(h(s))$.
Then,
\begin{align}
    \lefteqn{\int_{a}^{b} \Psi'(h(s)) \, \diffd s = \int_{k_3^2\Psi(h(a))}^{k_3^2\Psi(h(b))} \frac{k_3^{-2}\diffd x}{h'\comp h^{-1} \comp \Psi^{-1}(k_3^{-2}x)} }\hspace{0mm} \nonumber\\
&\le \int_{k_3^2\Psi(h(a))}^{k_3^2\Psi(h(b))} \frac{k_3^{-2}\diffd x}{\alpha \Psi^{-1}(k_3^{-2}x)} = \int_{k_3^2 \Phi_{k_3}^{-1}(h(a))}^{k_3^2 \Phi_{k_3}^{-1}(h(b))} \frac{k_3^{-2} \diffd x}{\alpha \Phi_{k_3}(k_3^{-2}x)} \nonumber \\
    &= \int_{\Phi^{-1}(k_3 h(a))}^{\Phi^{-1}(k_3 h(b))} \frac{\diffd x}{\alpha k_3 \Phi(x)}.
\end{align}
This establishes the result.
\QED

\bibliographystyle{abbrvnat}
\bibliography{sosm}

\end{document}